\documentclass[%
 aip,graphicx,twocolumn,amsmath,amssymb,
 reprint
]{revtex4-1}
\usepackage{graphicx}
\usepackage{dcolumn}
\usepackage{bm}
\usepackage{xcolor}

\begin{document}

\preprint{AIP/123-QED}

\title[]{Role of time scales and topology on the dynamics of complex networks }

\author{Kajari Gupta}
 \affiliation{Indian Institute of Science Education and Research (IISER) Pune, Pune-411008, India}
\author{G. Ambika}%
 \email{g.ambika@iisertirupati.ac.in}

\affiliation{Indian Institute of Science Education and Research (IISER) Pune, Pune-411008, India}
\affiliation{ 
Indian Institute of Science Education and Research (IISER) Tirupati, Tirupati-517507, India
}%

\date{\today}

\begin{abstract}
The interplay between time scales and structural properties of complex networks of nonlinear oscillators can generate many interesting phenomena, like amplitude death, cluster synchronization, frequency synchronization etc. We study the emergence of such phenomena and their transitions by considering a complex network of dynamical systems in which a fraction of systems evolves on a slower time scale on the network. We report the transition to amplitude death for the whole network and the scaling near the transitions as the connectivity pattern changes. We also discuss the suppression and recovery of oscillations and the cross over behavior as the number of slow systems increases. By considering a scale free network of systems with multiple time scales, we study the role of heterogeneity in link structure on dynamical properties and the consequent critical behaviors. In this case with hubs made slow, our main results are the escape time statistics for loss of complete synchrony as the slowness spreads on the network and the self-organization of the whole network to a new frequency synchronized state.
Our results have potential applications in biological, physical, and engineering networks consisting of heterogeneous oscillators. 
\end{abstract}

\maketitle

\begin{quotation}
Complexity of real world systems are studied mainly in terms of the nonlinearity in the intrinsic dynamics of their sub systems and the complex interaction patterns among them. In such systems, the variability and heterogeneity of the interacting sub systems can add a further level of complexity. In this context heterogeneity arising from differing dynamical time scales offers several challenges and has applications in diverse fields, ranging from biology, economy, sociology to physics and engineering. In our work, we study the interesting cooperative dynamics in interacting nonlinear systems of differing time scales using the frame work of complex networks. 
\end{quotation}
\section{\label{sec:level1}Introduction}

Multiple-timescale phenomena are ubiquitous in Nature and their in-depth understanding brings in several novel challenges.  Some of the examples of such phenomena in real world systems are neuronal electrical activity\cite{ogwa00,bullmore09,boyden05}, hormonal regulation\cite{gilman84,rad92,gold02,rivera00}, chemical reactions\cite{das13,andrews04,carter89}, turbulent flows\cite{scies87,hanja80,kim90} and population dynamics\cite{honey07,persson08,henzler07} etc.
Although there have been isolated studies addressing its various aspects, there are still many interesting questions that demand multidisciplinary approaches. Several modelling frameworks have proposed methods to understand dual time scale phenomena in single systems, like dynamical models for neuronal dynamics\cite{gilboa05}. However, studies on collective behavior of connected systems that differ in their intrinsic time scales, are very minimal with many open questions.  In this context, the framework of complex networks provides a promising tool to study nonlinear multiple time scale dynamics.

The emergence of synchronization in interacting dynamical units is important for the functionality of many systems and coupled oscillator networks are often studied to understand their dynamics\cite{alex08,stefano02}. Several types of synchronization phenomena like complete, phase and generalized synchronization\cite{meng03,jianquan05,batista07}, as well as cluster synchronization\cite{sarika16,sarika14} have been studied in various contexts.  However, frequency synchronization is of recent interest and has relevance in  many realistic situations ranging from neuronal systems to power grids\cite{motter13,levy00,fries01,janos01}, where the individual oscillators can have non-identical natural frequencies.

The suppression of oscillations or amplitude death\cite{renato,garima,awa10,ramana98,rajat07,resmi11,resmi12} is also an emergent phenomenon that has interesting implications. In an assembly of coupled systems, amplitude death emerges mainly due to specific nature of coupling like transmission delay, processing delay, dynamic coupling, nonlinear coupling, environmental coupling, etc.  or due to parameter mismatch \cite{rajat07,zhang11,zhao18,resmi11,sharma12,kamal15,sharmapre12,banerjee14,ghosh14,banerjee18} or time scale diversity\cite{silvia03,wei18}. Most often, amplitude death is required to suppress unwanted oscillations in connected systems. 

In this context we note that the revival of oscillations is an equally important and related emergent phenomena in coupled oscillators. Often in many systems recovering oscillations from suppressed state is necessary for  their proper functioning, to maintain the output intensity in arrays of power generators, to get maximum output from coupled laser systems even with transmission delay, to sustain oscillations in interacting cardiac cells etc. \cite{motter13,menck14,jenkins13,tchu14}. Recently a few mechanisms to revive oscillations from quenched state in coupled systems have been reported\cite{morino13,maj13,zou10,zou13,wei15}.

Interestingly, heterogeneity of interacting systems plays an important role in the diversity and organization in many complex systems\cite{neill98,goh75,grimm00,cameron07}. The dynamical processes underlying their complexity often display phase transitions and analysis and prediction of such transitions are crucial for their proper functioning. Heterogeneity can arise from the pattern of interactions or nature of connectivity among the systems. Thus, in scale-free networks, the heterogeneity is mostly from the broad distribution of node degrees such that there exists a few nodes with very high degrees. In addition, heterogeneity due to different dynamical time scales is prevalent in many complex systems. It is shown that functional hierarchy emerges through a form of self-organization of two distinct types of neurons, with multiple timescales and hence not only the spatial connections between neurons but also the timescales of neural activity are important in  the mechanisms leading to functional hierarchy in neural systems\cite{yamashita08}. So also, widely different time scales are common in systems of chemical reactions. In a recent study in biochemical networks, the fastest time scales is shown to correspond to the chemical equilibrium between metabolites while the slower ones relate to more physiologically relevant transformations\cite{jamshidi08}. This brings out the need to analyse correlations between metabolites considering the characteristic time scales of the network. The heterogeneity in network connectivity is shown to result in frequency synchronized clusters in non-identical phase oscillators\cite{michael16}.

The main motivation of the present study is to analyze the onset and characterization of interesting collective dynamics or emergent behaviour in a network of connected systems with differing dynamical time scales. Thus the heterogeneity in dynamical time scales that we introduce is different from earlier studies where connected systems with nonidentical intrinsic frequencies are considered. In two coupled slow and fast systems and in minimal networks of three or four systems, onset amplitude death and occurrence of frequency synchronized states with tunable emergent frequencies are studied recently\cite{kaj16,kaj17}. By considering two types of complex networks, we study the interplay between heterogeneity in link structure or connectivity among the systems and their differing time scales that can lead to interesting collective behaviour. Our study primarily uses standard periodic oscillators of R{\"o}ssler and Landau Stuart type so that it brings out the amplitude variations and their cross over behaviours as the slowness factor increases.  

We find the difference in time scales and the heterogeneity in connectivity together can drive the whole network to frequency synchronized clusters. Increasing the heterogeneity in time scales by increasing the number of slow systems or the mismatch in time scales, the whole network settles to a state of no oscillations. The transitions to that state as well as recovery to slower oscillations with cross over in amplitudes are some of the interesting results of the study. 

We also address the important question of what happens if part or even one node of a scale free network of systems suddenly slows down and then how does it affect the performance of the whole network. In this case, the robustness of the network to such changes is studied in terms of the time taken for each node to escape from the synchronized state leading the whole network to desynchronized dynamics. We find this phenomenon of loss of synchrony settles in a time that decreases with the degree of the node that becomes slow first.  The desynchronization transition that happens over a characteristic time is followed by the reorganisation of the whole network to a frequency synchronized state and this self-organization time depends on the difference in time scales. In addition, we study the transition to amplitude death on a scale free network, with multiple time scales, drawn from a normal distribution and with time scales decided by the connectivity of the systems.

\section{Random networks of slow and fast periodic systems}
We construct a random network of N nodes where each node represents an n dimensional dynamical system. The dynamical equation of each node is taken as 
\begin{eqnarray}
\dot{X_i} = \tau_i F(X_i)+G \epsilon\tau_i\displaystyle\sum_{j=1}^N A_{ij}(X_j-X_i)
\label{neteqn}
\end{eqnarray}
where i=1, 2, ..... N. $F(X_i)$ represents the intrinsic dynamics of each node. $ A_{ij}$ represents the adjacency matrix of connections in the network with its elements having values 1, if the nodes i and j are connected and zero otherwise. The topological connectivity of the random network is defined by a parameter $p$, where $p$ is the probability with which any two nodes of the network are connected. 
G is diag(1,0,....) since we consider diffusive coupling between the first variables of the systems.
The systems on the network are identical in their intrinsic dynamics except that they evolve at different time scales, $t_i$, which are scaled as $t_i=\tau_i t$ to get the equation\ref{neteqn}. Then $\tau_i$ will function as a parameter defining the dynamical time scale of evolution of the $i^{th}$ system. We vary $\tau_i$ in the range (0.1,1) to introduce time scale diversity in the network of systems.

Choosing periodic R{\"o}ssler system as nodal dynamics, the equations that govern the dynamics are 
\begin{eqnarray}
 \dot{x}_i&=& \tau_i(-y_i-z_i) +\tau_i \epsilon \displaystyle\sum_{j=1}^N A_{ij}(x_j-x_i)
\nonumber\\
\dot{y}_i&=&\tau_i(x_i+ay_i) 
 \nonumber\\
\dot{z}_i &=& \tau_i(b+z_i(x_i-c))
\label{ch4prosseqn}
\end{eqnarray}
With the parameters chosen as a=0.1, b=0.1 and c=4, the intrinsic nodal dynamics is periodic.  

First, we consider the case of dual time scales, such that in the network, out of N identical systems, $m$ evolve on a slower time scale. The subset of oscillators with slower time scale is taken as S. Thus $\tau_i=\tau$ if i belongs to the set S and $\tau_i=1$ for other nodes. 
Then the value of $\tau$ indicates the extend of mismatch in time scales between the two sets of oscillators, smaller values of $\tau$ corresponding to larger mismatch. 
We analyse how the slowness of $m$ of the systems can affect the dynamics of the whole network. For this the system of equations in eqn.(\ref{ch4prosseqn}) are integrated using Adams-Moulton-Bashforth method, with time step 0.01 for 100,000 times and the last 10,000 values of the x-variables are used for calculations in the study. 
 
\subsection{Suppression and recovery of oscillations} 

\begin{figure}[h]
\centering
\includegraphics [width = \columnwidth]{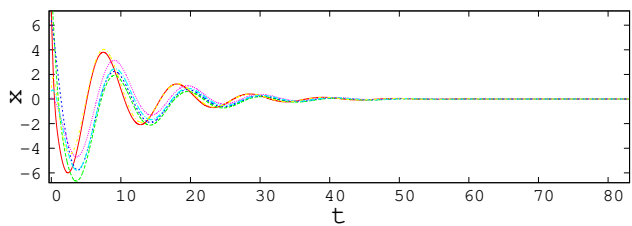}

\caption{\label{ch4randtime}(colour online) Time series of x variables of periodic R{\"o}ssler systems in a random network of slow and fast systems showing amplitude death state for $m$=50, $p=0.5$, $\tau$=0.35, $\epsilon=0.05$. }
\end{figure}

In this section, we report the general results on the coupled dynamics of the systems by varying the parameters involved, time scale of slow systems ($\tau$), number of slow systems ($m$) and coupling strength of connections($\epsilon$), keeping the probability of connection of the network $p=0.5$. We find that for sufficient time scale mismatch between slow and fast subsets of systems, for strong coupling, for a range $m$, all the systems go to a synchronized fixed point. This state is generally known as amplitude death(AD)\cite{renato,garima,awa10,ramana98,rajat07,resmi11,resmi12} in the context of coupled dynamics. The Fig.~\ref{ch4randtime} shows the state of amplitude death in the random network of slow and fast systems.

\begin{figure}[h]
\centering
\includegraphics [width = 0.6\columnwidth]{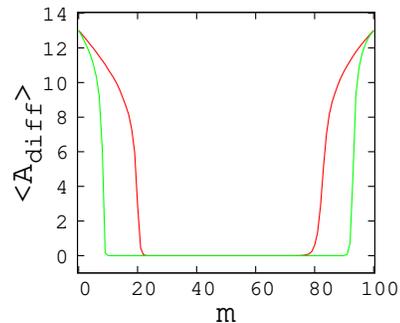}
\caption{\label{ch4amp_mbyn}(colour online)Variation of average $A_{diff}$ with $m$. Here $\tau$=0.35 and $\epsilon$=0.01 for red curve and $\tau$=0.35 and $\epsilon$=0.05 for green curve. N=100, $p$=0.5.}
\end{figure}

We calculate the difference between global maximum to global minimum, $A_{diff}$, from the time series of each oscillator. This averaged over all the N systems in the network serves as an index to identify onset of AD in the network, since $<A_{diff}>=0$ would correspond to AD state in the whole network. Using this we identify the region for occurrence of AD for different $m$, the number of slow systems present in the network, with suitable values chosen for the other parameters, $p$, $\tau$, $\epsilon$.  We plot this region for two sets of values of $\tau$ and $\epsilon$ in Fig.~\ref{ch4amp_mbyn}.
This shows that a minimum number of slow systems is required for AD to occur, denoted as $m_1$. As $m$ increases the heterogeneity decreases and the network recovers from AD state beyond a certain value, $m_2$. Thus, suppression of dynamics happens as $m$ reaches a critical minimum value $m_1$ and recovery to oscillatory state happens beyond the second critical value, $m_2$; both these values depend on other parameters like $\tau$, $p$ and $\epsilon$ of the system. 

\begin{figure}[h]
\centering
\includegraphics [width = 0.6\columnwidth]{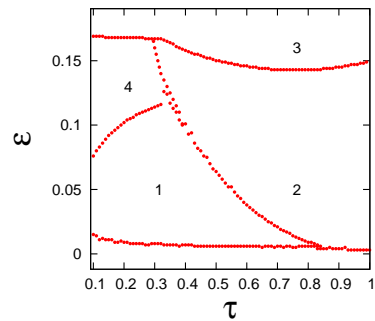}
\caption{\label{paraplanerandom}(colour online)Regions of different dynamical states in the parameter plane ($\tau, \epsilon$). Region 1 corresponds to AD, 2 corresponds to frequency synchronization, 3 leads to instability and 4 is a transient state where systems start diverging from fixed point state before reaching unstable state. Here $m$=50,$p$=0.5,N=100.}
\end{figure}

With $m$ chosen from the region of AD in Fig.~\ref{ch4amp_mbyn}, we isolate the region of AD in ($\tau, \epsilon$) plane, for a chosen $p=0.5$, and this is shown as region 1 in Fig.~\ref{paraplanerandom}.

\subsection{Frequency synchronized dynamics}
In this section we present the possible dynamical states outside the region of AD in the parameter plane ($\tau, \epsilon$). When coupling is strong and time scale mismatch is small, all the systems in the network settle to an organized state with oscillations of differing amplitudes but same frequency. This state of frequency synchronization is seen in region 2 in the Fig.~\ref{paraplanerandom}. This is identified by calculating the frequency of each oscillator from its x-time series using equation, 

\begin{eqnarray}
\omega =\frac {1}{K}\displaystyle\sum_{k=1}^{K}  \frac {2\pi}{(t_ {k+1}-t_k)}
\label{freq}
\end{eqnarray}
where $ t_k $ is the time of the $ k^{th} $ zero crossing point in the time series of the oscillator and K is the total number of intervals for which the zero crossings are counted.
In this state, the oscillations of slow systems are relatively closer in phase and so are fast oscillators among themselves but the phase difference between slow and fast sets is relatively large. 
Below region 2, with low coupling strength the oscillators show a two-frequency state and as time scale mismatch increases they become periodic with two separate time scales. For very high coupling strength in the region marked as 3, network becomes unstable but before this there is region for low $\tau $(region 4), where the systems are in a transient state, diverging from AD (Fig.~\ref{paraplanerandom}).

\subsection{Crossover phenomena in the emergent dynamics for large $m$}
\begin{figure}[h]
\centering
\includegraphics [width = 0.6\columnwidth]{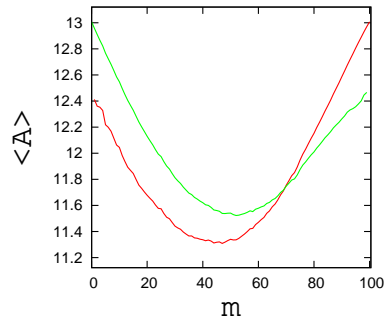}
\caption{\label{ch4crossover}(colour online)Crossover in magnitudes of amplitudes of slow and fast sets of oscillators, in the random network of 100 systems. Average amplitude of slow set is shown in (red) and that of fast in green. Here $p$=0.5, $\tau=0.7$, $\epsilon=0.05$.}

\end{figure}

When the systems are in the state of frequency synchronized oscillations, corresponding to region 2 in Fig.~\ref{paraplanerandom}, the amplitudes of slow and fast sets of systems vary from each other. In general, for low $m$, we observe that amplitudes of slow systems are smaller than those of fast systems, while for higher $m$ this behavior gets reversed with the slow set having larger amplitudes than the fast one. Thus, we observe a novel phenomenon of crossover behavior in the amplitudes as $m$ is varied. To show this explicitly, we study the average amplitude of all the slow systems and that of all the fast systems as $m$ is varied keeping the values of $\tau$ and $\epsilon$ in frequency synchronization state. We find that at a critical value of $m$, the amplitude of slow and fast systems undergoes a reversal as shown in Fig.~\ref{ch4crossover}.

When m=0 or N, the systems are all completely synchronized.  For other values of m,  there is approximate synchronization among systems in each set, slow and fast, but are phase shifted between them.  Then the number of systems included in the coupling that come with mismatch or the number of systems from the other set, is effectively deciding the amplitude of oscillations.  Thus for small m, the slow systems have smaller amplitude due to the larger number of phase shifted and mismatched systems from the fast set. This gets reversed as m becomes large and close to N. At a certain value of m, the cross over point, the effect of mismatched and phase shifted terms in the coupling have  equal effects for both sets.

\begin{figure}[h]
\centering
\includegraphics [width = 0.6\columnwidth]{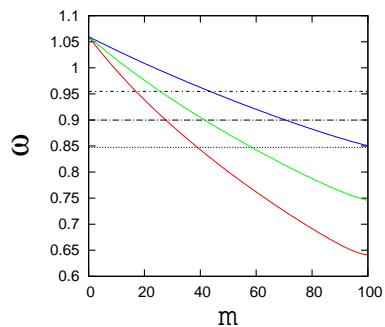}

\caption{\label{freqcrossrand}(colour online) Frequency vs $m$ for $\epsilon$=0.05 $\tau$=0.6(red),0.7(green) and 0.8(blue) showing crossover to frequency suppressed state as the emergent frequency crosses the mean value of fast and slow frequencies shown by corresponding black lines(increasing order of $\tau$ from below to above) . $p$=0.5.}
\end{figure}

In the state of frequency synchronization, we also observe a similar cross over in the synchronized frequency, which is high for low $m$ and very low for high $m$. The synchronized frequency calculated using eqn.(\ref{freq}) for all the oscillators is plotted as function of $m$ in Fig.~\ref{freqcrossrand}. For each particular $\tau$ the mean frequency of the intrinsic slow and fast frequencies is also shown (black lines). Thus, when the synchronized frequency for any given $\tau$ crosses the mean and decreases below that with larger $m$, we say frequency suppression occurs. The value of $m$ for which this happens is noted as the crossover point for the emergent frequency of the oscillators.

\subsection{Transition to amplitude death and connectivity of the network}
\begin{figure}[h]
\centering
\includegraphics [width = 0.4\columnwidth]{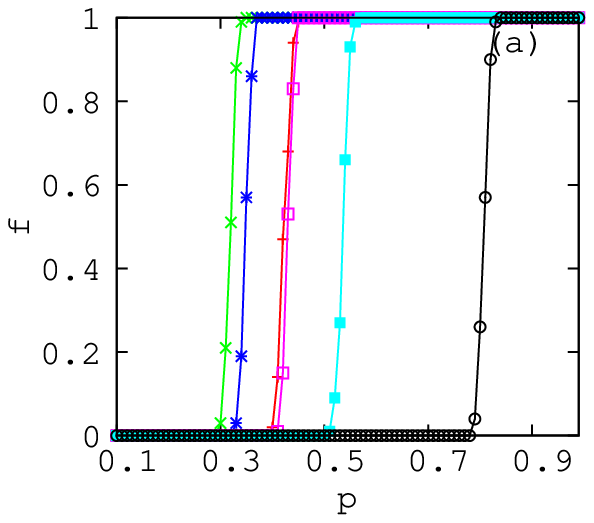}\includegraphics [width = 0.4\columnwidth]{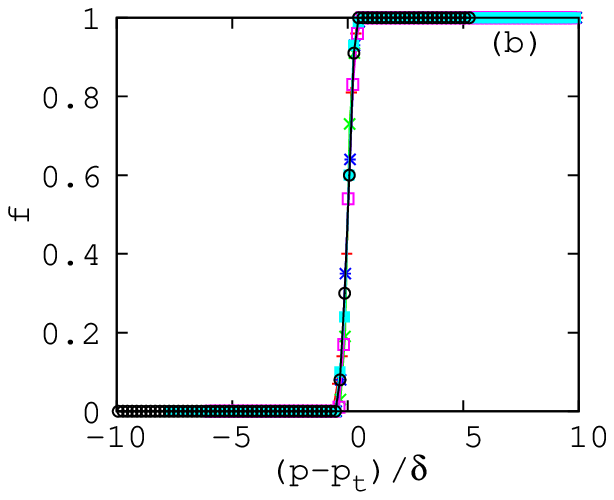}
\includegraphics [width = 0.4\columnwidth]{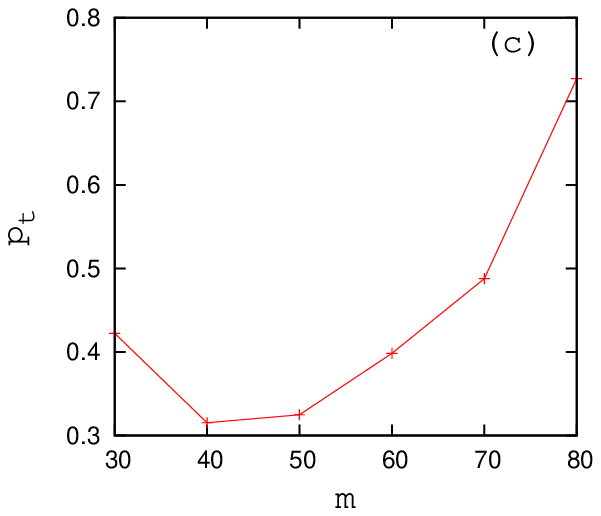}	

\caption{\label{ch4pr_trans}a)(colour online)a)Fraction of realizations $f$ for the transition to AD, plotted with the probability $p$ of connections in the network, b) normalized transition curves for $m$=30(red, plus), 40(green, cross), 50(blue, star), 60(magenta, square), 70(cyan, solid square), 80(black, circle) and c) Variation of critical $p$ value, $ p_t$ with $m$. Here $\tau=0.35$ and $\epsilon=0.01$.}
\end{figure}

The topology or structure of the random network used, depends on its connectivity which is decided by the probability of connections $p$. As $p$ is increased from 0 to 1, the topology goes from sparsely connected to fully connected network. In order to understand the role of topology or the connectivity of the network in the transition to AD, we study the collective behavior of all the systems by varying $p$, for an $m$ value that lies in the AD region of Fig.~\ref{ch4amp_mbyn} and values of $\tau$ and $\epsilon$ from the AD region in the parameter plane (Fig.\ref{paraplanerandom}). For each value of $p$, we take 100 realizations of the network and check what fraction of them goes to AD state for the whole network. This fraction of the realizations $f$ gives the probability of transitions. We plot $f$ for different values of $m$, to get the corresponding transition curves. We observe that as $m$ increases, the transition to AD occurs at lower values of $p$, till it reaches a minimum and with further increase of $m$, the transitions move to higher values of $p$. This is clear from Fig.\ref{ch4pr_trans}. 
The threshold value for the transition, where half of the realizations go to amplitude death, is taken as $p_t$. 

For each value of $m$ used, we get the width of the transition curve as $\delta$ and normalize the transition curves by replacing $p$ with $(p-p_t)/\delta$. Then we find all the transition curves fall on top of each other revealing a universal behavior. This data crunched curve is shown in the Fig.~\ref{ch4pr_trans}(b). Moreover, the threshold value $p_t$ varies with $m$ as shown in Fig.~\ref{ch4pr_trans}(c), which indicates a minimum at a particular value of $m$. At this value of $m$,  the heterogeneity in the network is optimized or maximized and hence transition to AD can happen even with sparsely connected network corresponding to minimum value of p, the probability of connections in the network. For the random network of 100 periodic R{\"o}ssler systems, $p_t$ is observed to be minimum when the number of slow systems is around 40. 

\subsubsection{Scaling with size of network}
\begin{figure}[h]
\centering
\includegraphics [width = 0.4\columnwidth]{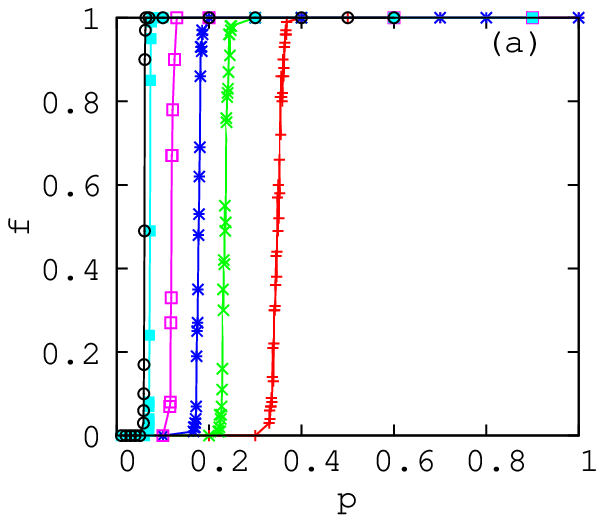}\includegraphics [width = 0.4\columnwidth]{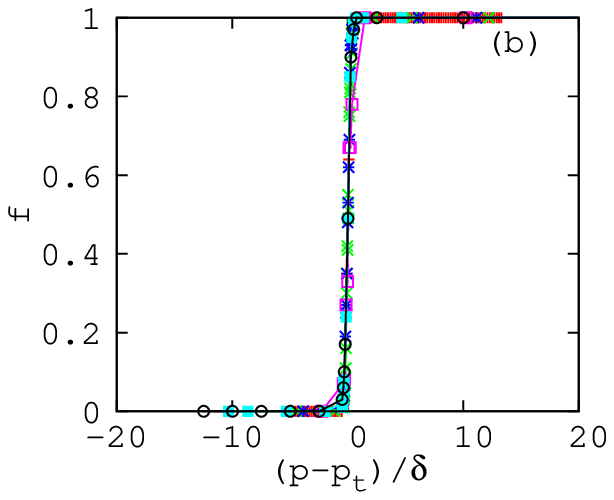}
\includegraphics [width = 0.4\columnwidth]{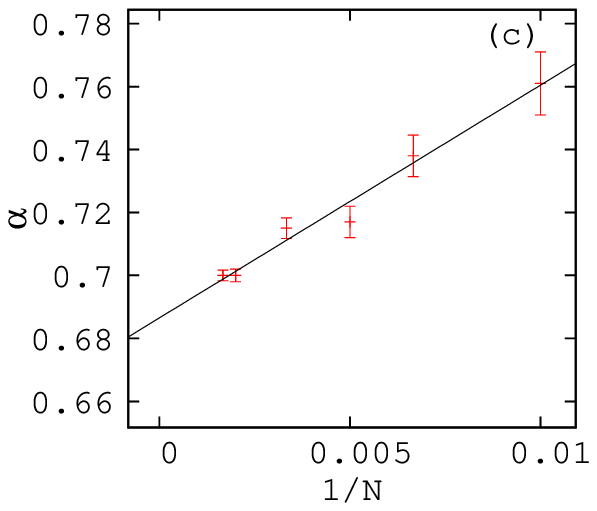}
\caption{\label{pr_trans_n}(colour online)Fraction of realizations $f$ for the transition to AD plotted with the probability $p$ for N=100(red, plus), 150(green, cross), 200(blue, star), 300(magenta, square), 500(cyan, solid square), 600(black, circle), $\tau=0.35$, $\epsilon=0.01$. b) Normalized transition curves $p-p_t/\delta$ for different N. c) Variation of the scaling index with 1/N with error bar shown in red.}
\end{figure}

We repeat the above analysis and obtain the transition curves for different network sizes, with N=100, 150, 200, 300, 500, 600, keeping $m$/N ratio fixed at 0.5. We notice the larger the size of the network lower the value of $p$ at which  transition takes place.  To study the scaling properties of these transitions, with system size, we fit each transition curve with the functional form 
\begin{equation} 
f=(p-p_c)^\alpha. 
\end{equation}
Here the value of $p_c$ is chosen as the one where the function gives best fit, and then the corresponding value of the scaling index $\alpha$ is calculated for each transition curve. Our results indicate that the index $\alpha$ varies with the network size N. To get the value of $\alpha$ in the large size limit, i.e. as N approaches infinity, we plot the calculated $\alpha$ vs 1/N and take the asymptote as 1/N goes to zero. This comes out to be 0.68, which within numerical errors, can be taken as 2/3. (Fig.~\ref{pr_trans_n})

\subsection{Transitions to amplitude death in random networks with non-uniform probabilities of connections}
In the study presented in sections IIA-IID, the probability of connections for generating the random network $p$ is kept the same for slow and fast nodes. We now consider a much more heterogeneous case of random networks, generated with three different probabilities and study the effect of slow and fast dynamics on it. This is done by taking the probability with which a slow system connects to another slow system as $p_1$, while a fast system connects with another fast system with $p_2$ and a slow system connects with a fast system with $p_3$. We compute the fraction of realizations resulting in amplitude death in this random network of slow and fast systems by varying $p_3$ for different sets of values of $p_1$ and $p_2$. 

\begin{figure}[h]
\centering
\includegraphics [width = 0.5\columnwidth]{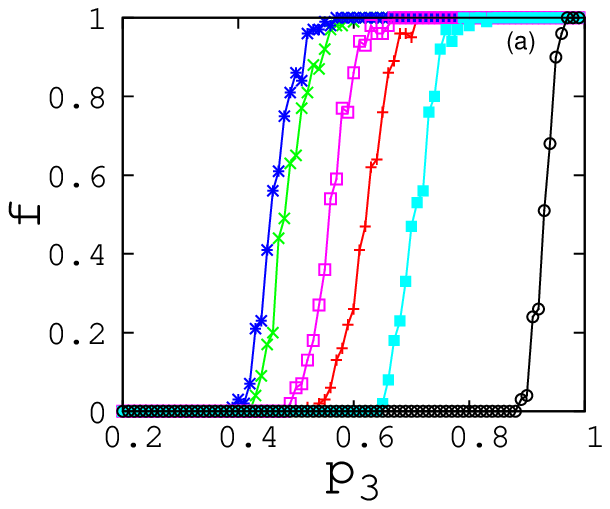}\includegraphics [width = 0.5\columnwidth]{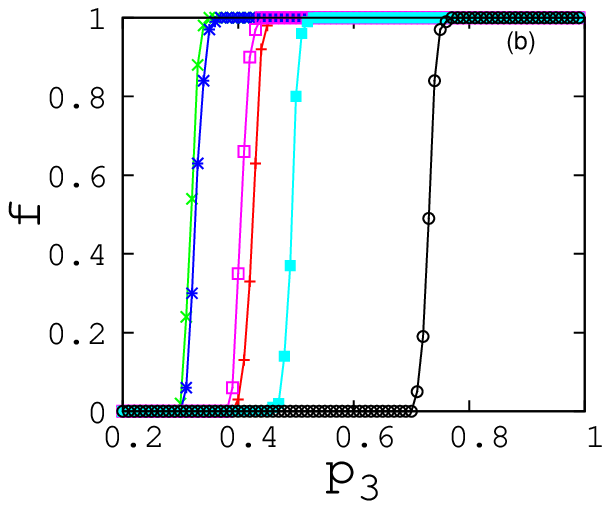}	
\caption{\label{pr3problowhigh}(colour online)Fraction of realizations for transition to AD for random network of heterogeneous probabilities for varying $p_3$ a)p1=0,p2=0, b)p1=0.3,p2=0.3. Here $\tau=0.35$, $\epsilon=0.01$. $m$=30(red, plus), 40(green, cross), 50(blue, star), 60(magenta, square), 70(cyan, solid square), 80(black, circle).}
\end{figure}

It is interesting to note that amplitude death happens even in a bipartite network, with $p_1=0$ and $p_2=0$ but non-zero $p_3$. However, having non-zero values for $p_1$ and $p_2$ helps the network to reach amplitude death state at lower values of probability $p_3$ and the minimum $p_3$ for this transition becomes smaller with increasing $p_1$ and $p_2$ (Fig.~\ref{pr3problowhigh}).  
\begin{figure}[h]
\includegraphics [width = 0.5\columnwidth]{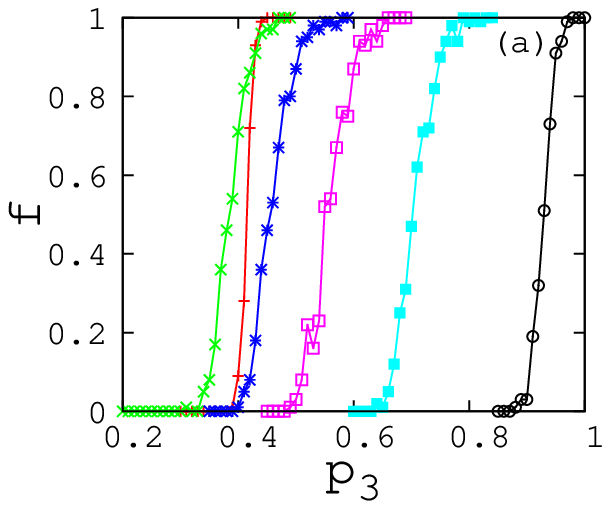}\includegraphics [width = 0.5\columnwidth]{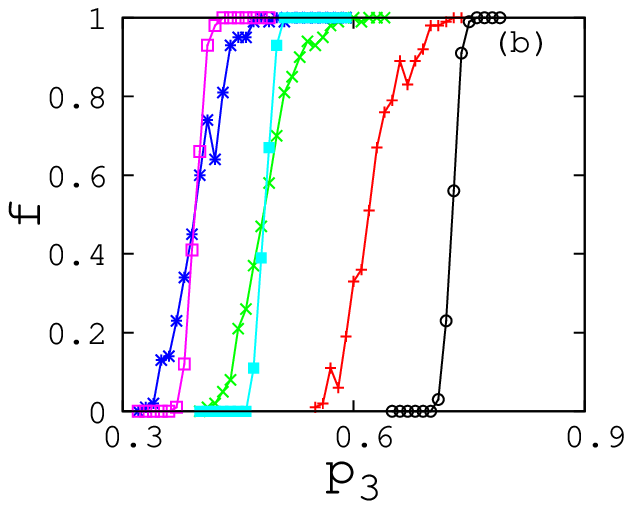}	

\caption{\label{prslowfastzero}(colour online)Fraction of realizations for transition to AD for a random network of heterogeneous probabilities for varying $p_3$ a)p1=0,p2=0.8, b)p1=0.8,p2=0, keeping $\tau=0.35$, $\epsilon=0.01$. $m$=30(red, plus), 40(green, cross), 50(blue, star), 60(magenta, square), 70(cyan, solid square), 80(black, circle).}
\end{figure}
We also study the special cases when with $p_1=0$, $p_3$ is varied keeping $p_2$ fixed as well as $p_2=0$ and $p_3$ is varied with $p_1$ fixed. The results shown are for $p_1=0.8$ and $p_2=0.8$ in the respective cases and $p_3$ is varied (Fig.~\ref{prslowfastzero}a,b).

\subsection{Random network of Landau Stuart systems}
\begin{figure}
\centering
\includegraphics [width = 0.48\columnwidth]{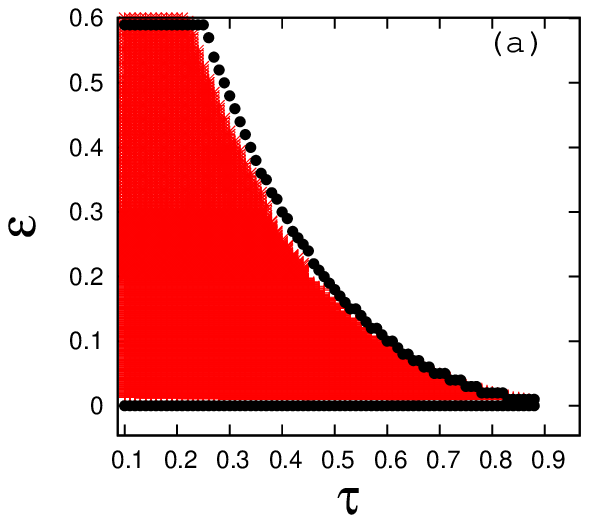}\includegraphics [width = 0.48\columnwidth]{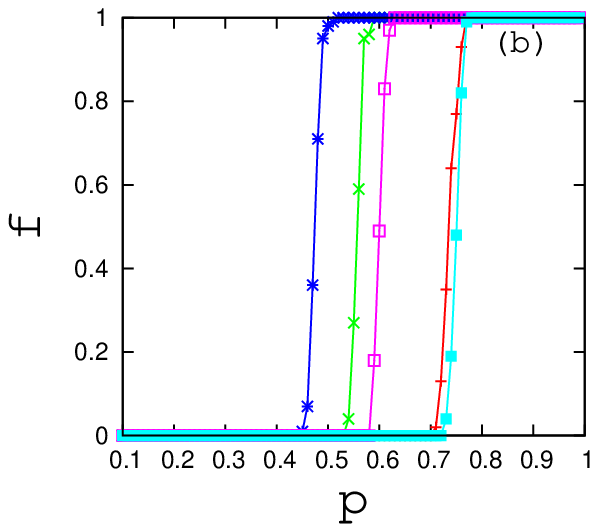}

\caption{\label{landaurand}(colour online)a)Region showing AD in $\tau,\epsilon$ plane for Landau-Stuart systems on a random network, $p=0.5$, $m=50$, $N=100$, b) variation of fraction of realizations with $p$ for $\tau=0.35$, $\epsilon=0.01$. $m$=30(red), 40(green), 50(blue), 60(magenta), 70(cyan).}
\end{figure}

We do a similar study for slow and fast Landau Stuart oscillators on random networks given by the equations
\begin{eqnarray}
 \dot{x}_i&=& \tau_i((a-{x_i}^{2}-{y_i}^2)x-\omega y_i)  +\tau_i \epsilon \displaystyle\sum_{j=1}^N A_{ij}(x_j-x_i)
\nonumber\\
\dot{y}_i&=&\tau_i((a-{x_i}^2-{y_i}^2)y+\omega x_i)
\label{ch3landaurandom}
\end{eqnarray}
The results are qualitatively similar for amplitude death and oscillatory behaviors. The region of AD is numerically calculated and shown for $p$=0.5, $m=50$, N=100 in $\tau,\epsilon$ (Fig.\ref{landaurand}a). 
Choosing the $\tau$ and $\epsilon$ from the region of AD one can show the variation for fraction of realizations of AD while $p$ varies. In this case also an optimum number of slow systems exists for which the transition occurs for minimum value of probability $p$ as shown in Fig.\ref{landaurand}b.

\subsubsection{Stability analysis for amplitude death state}

We find the onset of amplitude death is due to the stabilisation of the unstable fixed point at (0,0). Hence the transition to AD can be obtained by a detailed stability analysis. We note that stability of AD state is analysed in the case of two coupled systems of differing time scales, globally coupled networks and minimal networks or motifs of 3 or 4 systems\cite{silvia03,wei18,kaj16,kaj17}. In the present study, with heterogeneity in the pattern of connections also, AD occurs due to interplay of time scales and topology. 
 
We can write Jacobian of n-dimensional slow and fast systems on a network of size N as
\begin{equation}
{\bf J}= ({\boldsymbol \tau}{ \boldsymbol \cdot }{\bf I}){\boldsymbol \otimes}{\bf F'}+({\boldsymbol \tau}{\boldsymbol \cdot}{\bf L}){\boldsymbol \otimes}{\bf H}
\label{jacobian}
\end{equation}
where ${\boldsymbol \tau}$ is an NxN matrix in which $\tau_{ij}$ corresponds to $\tau_i$ of eqn(\ref{neteqn}) for all j. {\bf I} is NxN identity matrix. Dot product (${\boldsymbol \cdot}$) is defined here by the element wise product of two matrices, and cross product (${\boldsymbol \otimes}$) is defined as each element of the former matrix being multiplied by the later matrix as a block\cite{pecora}. {\bf L} is the laplacian matrix,
\begin{equation}
\bf L=\bf D-\bf A
\end{equation}
where {\bf D} is the diagonal degree matrix of the network, and {\bf A} is adjacency matrix.
${\bf F'}$ is Jacobian of the intrinsic system around fixed point and is an nxn matrix, where n is the dimension of a single system on each node. {\bf H} is nxn coupling matrix. 

In the case of Landau-Stuart oscillators systems with coupling function as given in eqn(\ref{ch3landaurandom}) and AD at the fixed point (0,0),  
we have
\begin{equation}
 {\bf F'} =   \left (\begin{array}{cc}
-a & -\omega   \\
\omega &  a   \end{array} \right), \nonumber
  {\bf H} =   \left (\begin{array}{cc}
\epsilon & 0   \\
0 & 0   \end{array} \right) 
\label{fandh}
\end{equation} 

We construct the matrix {\bf J} following the equation (\ref{jacobian}) and calculate its eigenvalues. Scanning the whole plane of $\tau$ $\epsilon$, we estimate the values of $\tau$ and $\epsilon$,  where the real part of the largest eigenvalue crosses zero. The plot of these values in ($\tau,\epsilon$) plane gives the transition curve to AD or the boundary for the stable region for AD. This is shown in black dots in Fig.\ref{landaurand}a. We note that the boundaries calculated using the above stability analysis agrees well those obtained by direct numericalGraph partitions and cluster synchronization in networks of oscillators simulations.

\section{Scale free networks of slow and fast systems}

As is well known, scale-free networks are inherently more heterogeneous than random networks, with broad distribution of nodal degrees and a few nodes with very high degrees, called hubs. Hence the emergent dynamics due to interactions among slow and fast dynamical systems on such a scale free network will be interesting. For this we generate several realizations of scale free networks using Barab{\'a}si-Albert algorithm\cite{albert2002statistical} and consider the dynamics on each node of the network, as that of periodic R{\"o}ssler systems. The equations for the dynamics on such a network will be the same as eqn(~\ref{ch4prosseqn}) with all parameters taken the same way. But the adjacency matrix $A_{ij}$ is taken as per the scale free network topology obtained from the Barab{\'a}si-Albert algorithm. 

\begin{figure}
\centering
\includegraphics [width =\columnwidth]{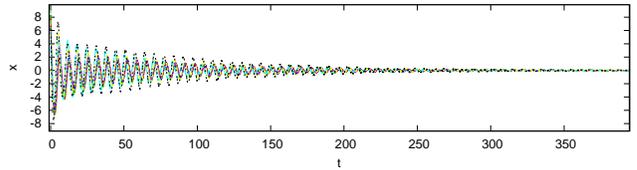}
\caption{\label{sf_ad}(colour online) Time series of few typical x-variables are plotted, showing amplitude death in a scale free network of periodic R{\"o}ssler systems for $\tau=0.3$, $\epsilon=0.2$ and $m$=8.}
\end{figure}

\begin{figure}
\centering
\includegraphics [width = 0.7\columnwidth]{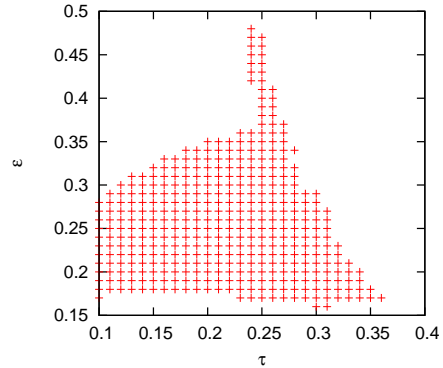}

\caption{\label{sf_ad_para}(colour online) AD region in ($\tau,\epsilon$) plane for 100 periodic R{\"o}ssler systems coupled on a typical realization of scale free network with 8 hubs taken from the high degree end slow in dynamics.}
\end{figure}

In a typical calculation, we take a network of size 100, with a set of $m$ nodes evolving at the slower time scale. Since in a scale free network, hubs play the role of control nodes, we mostly concentrate on cases where hubs follow slower dynamics. Hence in this case the number of slow systems required for AD to occur, is much smaller. Thus on a scale free network of 100 systems even with eight of the higher degree nodes or hubs having a time scale mismatch of $\tau$, we find the dynamics of all the systems can be suppressed to AD state (Fig.\ref{sf_ad}).
We isolate the region of AD in ($\tau,\epsilon$) plane as the region where the difference between the global maxima and global minima of all the oscillators goes to zero. This is shown in Fig.~\ref{sf_ad_para}. 

\begin{figure}
\centering
\includegraphics [width = 0.7\columnwidth]{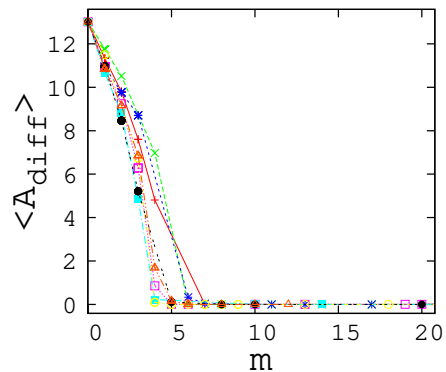}

\caption{\label{sf_ad_m}(colour online)$<A_{diff}>$ vs $m$ showing the minimum value for number of slow hubs required for AD to occur for periodic R{\"o}ssler systems on a scale free network. Here $\tau=0.12$ and $\epsilon=0.25$ Different colors correspond to different realizations.}
\end{figure}

We also estimate the minimum number of slow systems required to induce AD starting with the highest degree as slow and increasing the number one by one. For each case, the average amplitude differences of all the oscillators is calculated. The plot of this averaged amplitude $(<A_{diff}>)$ with the number of slow hubs $m$ gives this as the value of $m$ at which $(<A_{diff}>)$ becomes zero. This is repeated for different realizations and shown in Fig.\ref{sf_ad_m}. Here the values of $\tau$ and $\epsilon$ are chosen from the AD region in Fig.~\ref{sf_ad_para}.

We repeat the above study using Landau-Stuart oscillator as nodal dynamics. We find qualitatively similar results for this case also with AD state and frequency synchronization.

\subsection{Spreading of slowness on scale free networks}
When there is no time scale mismatch in the dynamics of systems, all the systems on a scale free network, can be completely synchronized with a sufficiently strong coupling strength. Starting with such a state, after giving sufficient time so that all the oscillators settle to complete synchronization, we make one of the nodes, called source node, slower in its dynamics. Clearly this can disrupt the dynamics of all other nodes as the slowness spreads over the network. Consequently, all oscillators will then move away from the state of complete synchronization.
\begin{figure}
\centering
\includegraphics [width = 0.9\columnwidth]{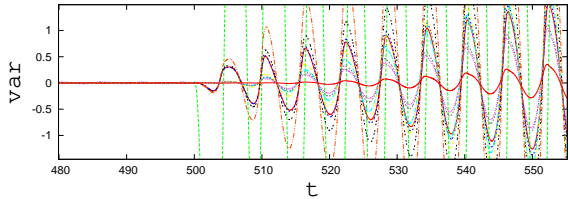}
\caption{\label{scalefreevar}(colour online)Variance of few typical periodic R{\"o}ssler oscillators (var) with time to show that each oscillator takes a different time to move away from the synchronized state. $\tau=0.3$, $\epsilon =0.03$.}
\end{figure}

\begin{figure}
\centering
\includegraphics [width = 0.45\columnwidth]{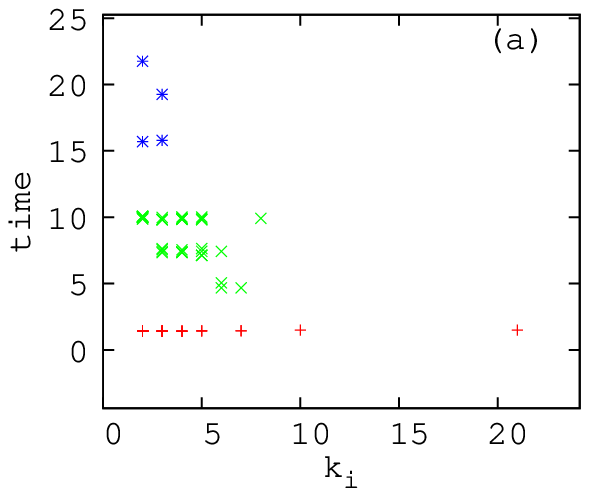}\includegraphics [width = 0.45\columnwidth]{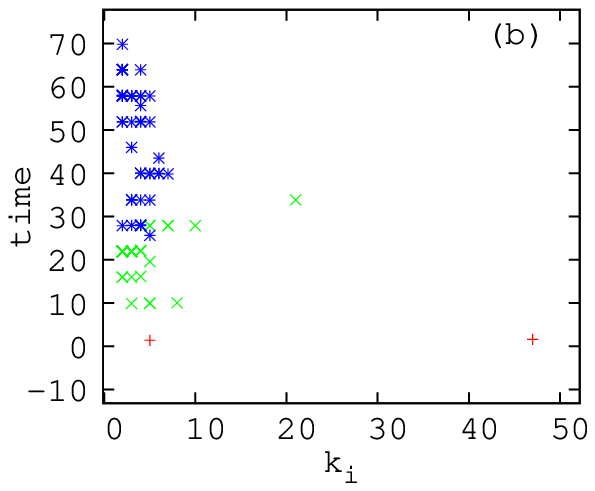}
\caption{\label{spreadofslowness1}(colour online)Time taken for each oscillator to move away from synchrony is plotted with its degree ($k_i$) when one source node becomes slow. This is shown for a particular realisation of the network of 100 systems where a) the highest degree hub is made slow with degree 47 and in b) the lowest degree node is made slow with degree 2. Different colors represent different shortest path lengths from the source node with shortest path 1(red, plus), 2(green, cross), 3(blue, star). Here $\tau$ is 0.3 for the source node and $\epsilon$ is taken as 0.03.}
\end{figure}

Due to the heterogeneity of connections in the scale free network, the time taken by each oscillator to move away from synchronization will not be the same. We analyze this scenario in terms of the degree of the node and shortest path from the source node, in the following two ways. We calculate the change in the variance of all oscillators in time. When they are completely synchronized, the variance would be zero as shown in Fig.~\ref{scalefreevar}. When one node is made slow, the variation of each oscillator from the mean of anticipated synchronized oscillations(the synchronized oscillation they would have followed if this node was not made slow), is nonzero indicating onset of desynchronization. From the Fig.~\ref{scalefreevar} it is evident that for each oscillator the time taken for the variance to go to a non zero value $ \pm \nu$ is different, with the source node taking the least time obviously.  This time, $t_\nu$, for each oscillator to reach a specific value $\pm \nu$ (typically -0.01 or 0.01) for its variation is plotted as a function of the degree of the nodes. It is easy to see that $t_\nu$ increases as the shortest path of that node from the source node increases. We repeat this for several nodes as sources, including hubs and low degree nodes. Fig.~\ref{spreadofslowness1} shows the plot of $t_\nu$ against degree of nodes for the two cases with a hub as the source node and a low degree node as the source node for a typical realization.  In the case where a hub is the source of slowness, we see most of the nodes move away in much shorter times since the shortest path from the source node is small. 
 
\begin{figure}
\centering
\includegraphics [width = 0.45\columnwidth]{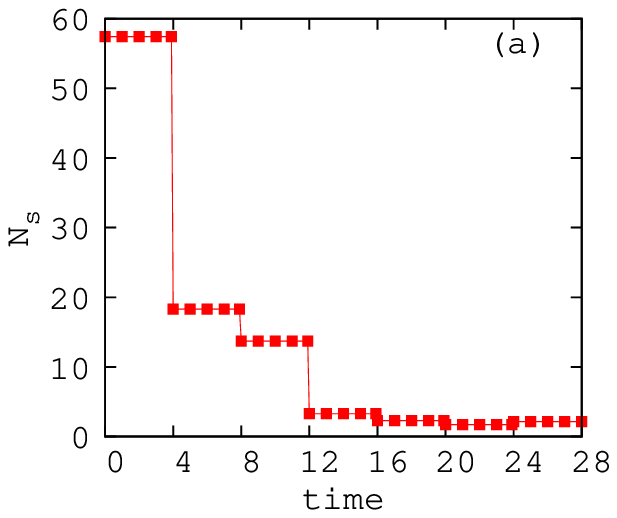}\includegraphics [width = 0.45\columnwidth]{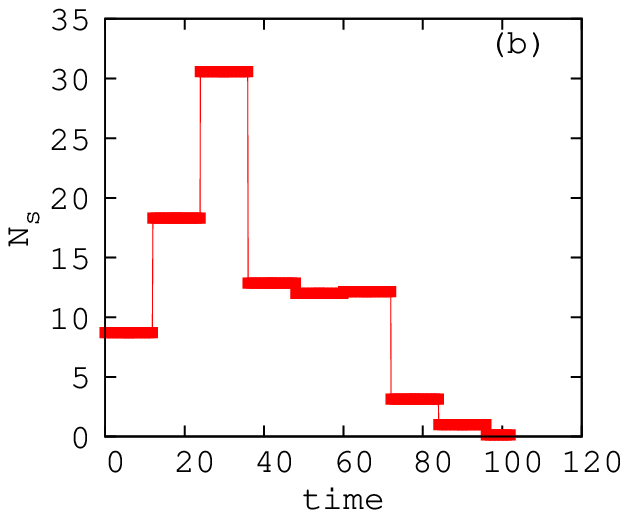}
\caption{\label{spreadofslowness2}(colour online) Number of systems that move away from synchrony($N_s$) in a range of time is plotted with time. The results shown are averaged over six realisations. Here a) corresponds to the highest hub as source of slowness for each realisation and b) corresponds to the lowest degree node being made slow. In this case also, $\tau$ is 0.3 for the source node and $\epsilon$= 0.03.}
\end{figure}

\begin{figure}
\centering
\includegraphics [width = 0.6\columnwidth]{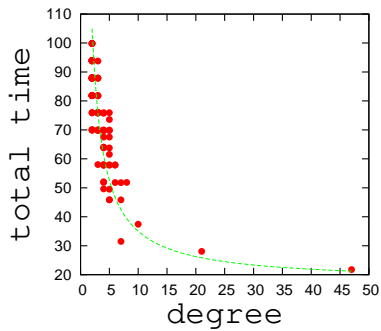}
\caption{\label{totaltimedegree}(colour online)Total time taken for all oscillators in the network to move away from synchrony is plotted against the degree of source node for periodic R{\"o}ssler systems on scale free network. The source node has time scale as $\tau=0.3$ and $\epsilon=0.03$.}
\end{figure}

We repeat the study for different realizations of the network and calculate the number of systems that get desynchronized in a given time and this number averaged over the realizations varies as shown in  Fig.~\ref{spreadofslowness2}, for the two cases, when a hub and a low degree node are the sources of slowness.
It is interesting to note that the total time taken for all the systems to move away from each other decreases with the degree of the source node as a power law.  This is clear from Fig.~\ref{totaltimedegree} shown for a typical realization of the network. We find that this curve can be fitted with a function as
\begin{equation}
T(k)=(a/k)+b
\label{fittimedegree}
\end{equation}
with $a=180$ and $b=18$. The desynchronization transition is followed by a reorganization of the whole network.
\subsection{Self organization of the network to frequency synchronized state}
\begin{figure}
\centering
\includegraphics [width = 0.9\columnwidth]{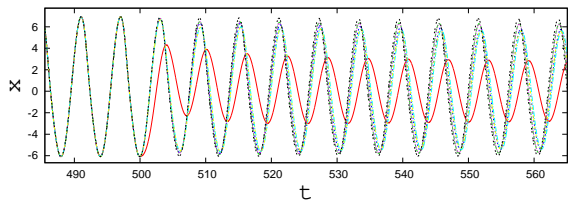}
\caption{\label{freqsyncscalefree}(colour online)Time series of few typical x-variables of periodic R{\"o}ssler systems are plotted to show the frequency synchronized state reached from the state of complete synchronization after one node is made slow. Here $\tau=0.3$ for the slow node and $\epsilon =0.1$.}
\end{figure}

Once synchrony is disturbed as discussed above due to a single node going slow, de-synchronization sets in characteristic times depending on the degrees of nodes in the network. Subsequent to this, given sufficient time, all the oscillators are found to reorganize themselves into a frequency synchronized state.  (Fig.~\ref{freqsyncscalefree}). This is an interesting and novel phenomenon of {\bf self-organization}, where the network goes from a collective behavior of complete synchronization to another less ordered but coherent emergent state of frequency synchronization by re-adjusting the dynamics of all the nodes, after the network is perturbed by making one node slow. 

\begin{figure}
\centering
\includegraphics [width = 0.5\columnwidth]{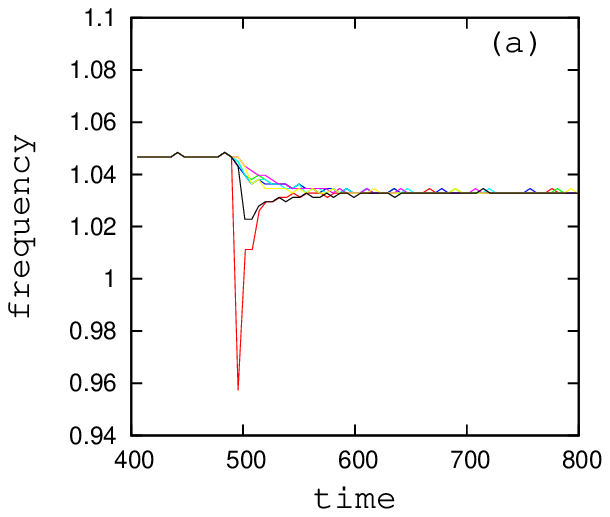}\includegraphics [width = 0.5\columnwidth]{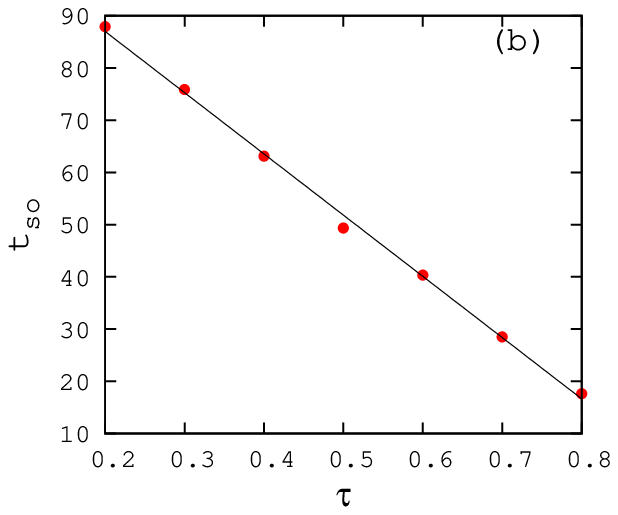}
\caption{\label{selforganization}(colour online)a) Self-organization of oscillators into frequency synchronized state for the whole network for a typical realization of scale free network after one hub is made slow. On the x axis the time and on the y axis the corresponding frequency are shown. At time=500 one hub is made slow, b) Time taken to organize into the new synchronized state, $t_{so}$, averaged over six realizations is plotted for different $\tau$ values with $\epsilon$ kept at $0.1$. $t_{so}$ falls off with increasing mismatch $\tau$ as a straight line.}
\end{figure}

To characterize this process, we calculate the frequency of each oscillator using eqn.\ref{freq} and plot them with time(Fig.\ref{selforganization}a). The figure shows the synchronized frequencies in the beginning, the de-synchronized frequencies just after one node is made slow at t=500, and finally the re-adjusted lower frequency after self-organization to frequency synchronized state. We also study the time taken for self-organization, called self organization time, ($t_{so}$), averaged over several realizations for $\tau$ varying in the range $0.2-0.8$ for a particular $\epsilon$=0.1, (Fig.\ref{selforganization}b). We observe that $t_{so}$ decreases with the increase in the time scale mismatch introduced on the source node, along a straight line. 
\subsection{Scale free network with multiple time scales}
In this section, we study the collective dynamics of nonlinear systems on a scale free network where the time scale of each node varies with its degree following the relation
\begin{equation}
\tau_i=2/k_i
\label{timescaledistribution}
\end{equation}

\begin{figure}
\centering
\includegraphics [width = 0.7\columnwidth]{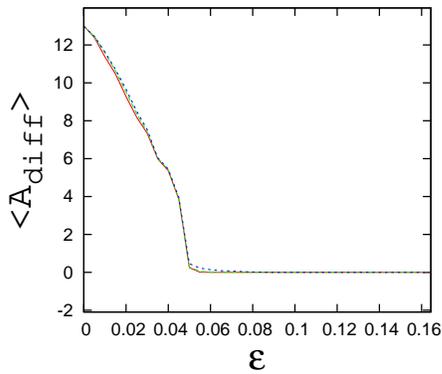}
\caption{\label{adinepsilon}(colour online)$<A_{diff}>$ vs $\epsilon$ in a typical realization of scale free network showing onset of AD due to multiplicity of time scales for periodic R{\"o}ssler systems with $\tau_i=2/k_i$ for N=100(red), 500(green), 1000(blue).}
\end{figure}

This is chosen such that the node with highest degree will have the slowest time scale and the time scale increases as degree decreases. Since the network is scale free, the number of different degrees, $k_i$  and hence the number of different time scales, $\tau_i$,  will be less than N but still will have a multiplicity of time scales. Thus, for one typical realisation of 1000 nodes, we get 30 different time scales in the network.  We find the presence of multiple time scales, forces the whole network to collapse to a state of AD. The onset of AD, after a threshold coupling strength, is evident from the plot of $<A_{diff}>$, for different values of $\epsilon$, in Fig.~\ref{adinepsilon} with three different sizes N=100,500 and 1000. 

\begin{figure}
\centering
\includegraphics [width = 0.7\columnwidth]{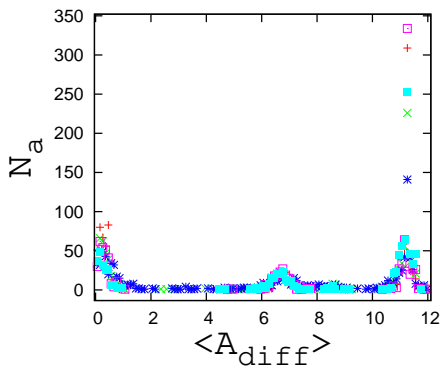}
\caption{\label{amptaudistribution}(colour online)Distribution of amplitudes of 1000 periodic R{\"o}ssler systems with multiple time scales on a scale free network for $\epsilon=0.03$, with the time scales distributed as eqn.~\ref{timescaledistribution}. The different colours indicate different realizations of the network.}
\end{figure}

For lower values of $\epsilon$, prior to onset of AD, we see oscillations with differing amplitudes. But even with multiple time scales, the connectivity through the network makes the systems organize into three groups, high degree nodes with lower time scales having smaller amplitudes, low degree nodes with faster time scales having larger amplitudes and an intermediate group with amplitudes in between. This is clear from the distribution of amplitudes of all the oscillators in the network from six realizations of the network of size N=1000 (Fig.\ref{amptaudistribution}). 

\begin{figure}
\centering
\includegraphics [width = 0.5\columnwidth]{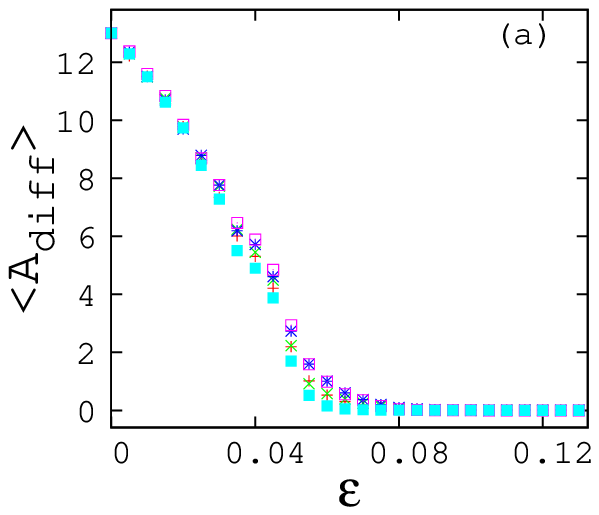}\includegraphics [width = 0.5\columnwidth]{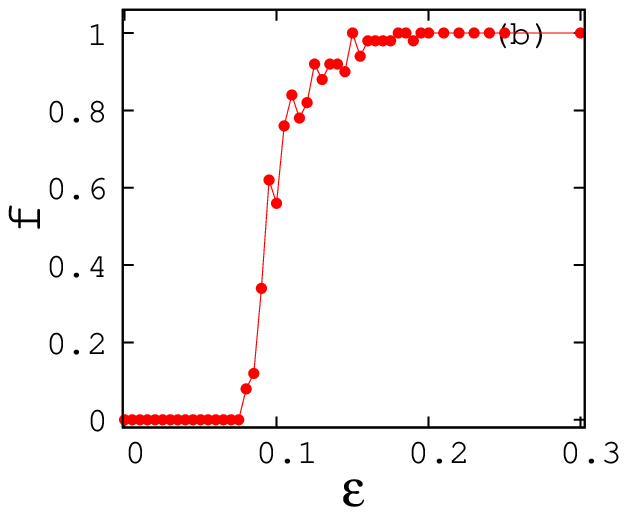}
\caption{\label{normaltau}(colour online)Onset of AD on a scale free network of 100 periodic R{\"o}ssler systems with normal distribution for time scales. a)$<A_{diff}>$ vs $\epsilon$ in 5 typical realizations (shown in different colour) of time scale distributions showing onset of AD. b)Fraction of realizations of time scale distributions showing transition to AD with $\epsilon$  for 50 realizations. }
\end{figure}

Similarly we do the analysis for another distribution of time scales on nodal dynamics of scale free network. In this case we choose the $i^{th}$ node to follow a time scale $\tau_i$ drawn from a normal distribution with mean $0.5$ and standard deviation $0.15$. Since in this case there is no correlation between the degree of the node and its time scales and the network is hierarchical, we take multiple realizations of the distribution and analyse the dynamics. The onset of AD state on a scale free network of periodic R{\"o}ssler systems and the threshold value of $\epsilon$ is shown in (Fig.~\ref{normaltau}a). We also estimate the transition probability as the fraction of realizations of time scale distributions for which AD occurs, for different values of $\epsilon$ and plot this fraction (f) with $\epsilon$ to see the transition curve to AD. (Fig.~\ref{normaltau}b) 
\section{Conclusion}
The study reported, in general, addresses the important question of how the mismatch in dynamical time scales of different interacting units can affect the collective performance of a complex system. We consider nonlinear dynamical systems with complex interaction patterns modelled by random and scale free topologies. We are motivated by the fact that one important factor for heterogeneity of interacting systems in real-world complex systems is the diversity in time scales. This makes our study highly relevant in understanding such systems from a dynamical system's perspective. 

We study the robustness of coupled oscillator networks which are widely used as models for understanding the dynamics of networked systems in biology, physics, and engineering and see how increasing the heterogeneity in time scales makes them undergo transitions in their dynamics. The emergent dynamics is then characterized in terms of the average amplitude of oscillations and the common frequency of the coupled systems. 

The results presented are primarily for periodic R{\"o}ssler systems, with two distinct time scales. However qualitatively similar results are obtained in the case of random networks, with Landau-Stuart oscillators, chaotic R{\"o}ssler and Lorenz systems.  

When the network of connections is random, we study suppression and recovery of oscillations and cross over in amplitudes and frequency as number of slow systems increases. The transition to AD in terms of the probability of connections p is found to scale with network size, the index of scaling being 2/3. 

If the systems are connected on a scale free network with heterogeneity in the degrees, we find hubs can function as control nodes in the emergent dynamics. We study the spread of slowness through the network due to one hub being slow and discuss the self-organization of the whole network from completely synchronized state to a state of frequency synchronization. The hierarchical structure of this network leads to different patterns in the spread of slowness due to one slow source node. The study of the times involved gives an estimate of minimum time within which corrective measures are to be initiated to restore the network dynamics. 

We also observe transition to AD on a scale free network, with multiple time scales, with time scale of each node depending on its degree and with time scales drawn from a normal distribution.  

In all cases studied, suppression of dynamics with the whole network settling to a common fixed point, seems to be the most prevalent emergent state. We note that amplitude death due to time scale diversity has been reported in a few specific cases with global coupling\cite{silvia03,wei18}. Our study establishes how difference in dynamical time scales of systems induces AD in systems, interacting with random and scale free patterns of connections. So also we show recovery of oscillations of any desired common frequency is possible by tuning the time scales of the systems. It is interesting that cooperative phenomenon of lower order like frequency synchronization, may still emerge even in the presence of time scale diversity. As reported recently in modular networks of neurons\cite{kunal19}, tuning the time scales can also result in a variety of specific activity patterns with spatial order and frequency locking.


\end{document}